\documentclass[conference]{IEEEtran}
\IEEEoverridecommandlockouts
\usepackage{cite}
\usepackage{booktabs}
\usepackage{url}
\usepackage{amsmath, amssymb, amsthm, bm}
\usepackage{algorithm}
\usepackage{algorithmic}
\usepackage{graphicx}
\usepackage{textcomp}
\usepackage{xcolor}
\usepackage{hyperref}
\usepackage{orcidlink}
\def\BibTeX{{\rm B\kern-.05em{\sc i\kern-.025em b}\kern-.08em
    T\kern-.1667em\lower.7ex\hbox{E}\kern-.125emX}}
\begin{document}

\title{COAST: Congestion-Aware Start-Time Recommendations for Carbon-Aware HPC Jobs
\thanks{This work was supported by the NetDRIVE Flexible Fund under the UKRI Digital Research Infrastructure programme [UKRI910]. W. Feng was also supported by the UKRI EPSRC Doctoral Training Partnership [EP/W524414/1]. We thank Dr. Daniel Traynor and Prof. Jesus Lizana for their support and acknowledge the use of the Isambard 3 Tier-2 HPC Facility, funded by UKRI and EPSRC [EP/X039137/1].}}

\author{
\IEEEauthorblockN{
Weibin Feng$^{1}$\,\orcidlink{0009-0001-2441-5441},
Abhishek Dasgupta$^{2}$\,\orcidlink{0000-0003-4420-0656},
Zeynep Duygu Tekler$^{3}$\,\orcidlink{0000-0002-1858-0846},
Sudha Ahuja$^{4}$\,\orcidlink{0000-0003-4368-9285},
Jin Zheng$^{1}$\,\orcidlink{0000-0002-1783-1375},
Xun Jiang$^{5}$\,\orcidlink{0000-0001-5062-969X}
}
\vspace{6pt}
\IEEEauthorblockA{
$^{1}$\textit{School of Engineering Mathematics and Technology, University of Bristol}, Bristol, UK\\
$^{2}$\textit{Oxford Research Software Engineering, MPLS Doctoral Training Centre, University of Oxford}, Oxford, UK\\
$^{3}$\textit{Department of Engineering Science, University of Oxford}, Oxford, UK\\
$^{4}$\textit{School of Physical and Chemical Sciences, Queen Mary University of London}, London, UK\\
$^{5}$\textit{Department of Electrical and Electronic Engineering, Cardiff University}, Cardiff, UK
}
}

\maketitle

\begin{abstract}
High-performance computing (HPC) workloads consume substantial amounts of electricity, and their carbon emissions vary over time with the carbon intensity of grid electricity. However, uncoordinated shifting of carbon-aware HPC jobs toward low-carbon periods can concentrate recommended start times in the same time slots, creating congestion and eroding the resulting carbon benefits. This paper proposes COAST, a congestion-aware start-time recommendation mechanism that coordinates HPC jobs by balancing carbon savings against additional job delay and congestion in recommended start-time slots. COAST formulates each decision batch as an exact potential game, enabling best-response updates to converge to stable start-time recommendations. Under an idealized realization model that assumes each job can start at its recommended time, we conduct trace-driven simulations using public HPC energy data and historical grid carbon-intensity traces. The results show that COAST reduces estimated carbon emissions by 15.4\% with 1.5 hours of additional average start delay. Compared with carbon-greedy start-time recommendations, COAST reduces peak time-slot load by 23.8\% while preserving 93\% of the achievable carbon savings. These results quantify the potential benefits of congestion-aware start-time coordination for flexible, carbon-aware HPC jobs.
\end{abstract}


\section{Introduction}
\label{sec:introduction}

The carbon emissions associated with computing depend not only on the
energy consumed by workloads, but also on the carbon intensity of the
electricity used to power them. Since grid carbon intensity varies over
time, temporally flexible workloads can potentially reduce their
operational emissions by shifting execution toward lower-carbon
periods. Carbon-aware workload shifting has therefore emerged as a
promising direction for reducing the environmental impact of
large-scale computing systems \cite{9770383}. This
opportunity is particularly relevant to HPC, where jobs can be energy intensive and many workloads have some
flexibility in their execution time.

However, independently recommending the lowest-carbon period to every
flexible job can create a new coordination problem. If many users shift
their jobs toward the same favorable period, the resulting concentration
of resource demand may increase congestion and reduce the practical
benefit of carbon-aware timing. Here, congestion refers to the temporal
concentration of advised resource demand, modeled as a soft penalty rather
than scheduler queueing or a hard capacity constraint. The problem is not
simply to identify the cleanest time slot for an individual job; it is to
coordinate multiple jobs whose timing decisions jointly determine the
load profile. This issue is especially important when jobs have different
runtimes and resource demands, since jobs that begin at different times
may still overlap substantially during execution. Recent work has also highlighted that the benefits of carbon-aware
workload shifting can be sensitive to operational constraints and to
the assumptions used in evaluation \cite{10.1145/3627703.3650079}.

This paper proposes COAST, a congestion-aware
timing advice mechanism for flexible HPC jobs. COAST collects job
intents in configurable micro-batches and recommends target execution
start times over a future horizon discretized into configurable time
slots. Rather than optimizing each job in isolation, COAST balances three
factors:
predicted carbon emissions, user waiting time, and the additional
execution-slot congestion induced by assigning a job to a candidate
start time. We model the within-batch coordination problem as a finite
congestion game in which jobs select feasible target start times and
their costs depend on overlapping resource demand throughout their
execution intervals. The resulting formulation admits an exact
potential function, enabling COAST to compute stable recommendations
through sequential strict best-response updates. COAST is designed as an external advisory layer and does not replace the underlying HPC scheduler. We therefore evaluate the algorithmic potential of coordinated target start-time selection under an idealized start-time realization model.

The contributions of this work are as follows:
\begin{itemize}
\item We formalize the temporal concentration problem in carbon-aware
HPC workload shifting. In particular, we model congestion over
overlapping execution intervals, rather than treating carbon-aware
timing as a collection of independent per-job decisions.
\item We develop COAST, a micro-batch coordination mechanism that
internalizes each job's marginal contribution to execution-slot
congestion while balancing estimated carbon emissions and waiting
time. For each fixed batch, the resulting formulation is an exact
potential game, and sequential strict best-response updates converge
in finitely many steps to a stable assignment.
\item Through trace-driven simulation with public job-level energy
data and historical carbon intensity traces, we quantify the
carbon--delay--load trade-off of congestion-aware coordination
relative to observed and carbon-greedy timing baselines. COAST
attains most of the carbon reduction of carbon-greedy timing while
avoiding the load peaks that purely carbon-driven shifting creates,
lowering peak execution-slot load rather than raising it.

\end{itemize}

\section{Related Work}

Temporal workload shifting has become a central approach in carbon-aware computing. Radovanovi\'c et al. demonstrated its feasibility at datacenter scale through Google's Carbon-Intelligent Computing Management system, which uses carbon intensity forecasts and flexible computing capacity to defer workloads from less favorable periods \cite{9770383}. Wiesner et al. subsequently examined temporal shifting across workload durations, delay windows, regions, and forecast conditions, showing that its effectiveness depends strongly on both workload flexibility and the temporal characteristics of the local electricity grid \cite{10.1145/3464298.3493399}. More broadly, Sukprasert et al. quantified the ideal and practical limits of carbon-aware temporal and spatial shifting across diverse cloud workloads, finding that deadlines, service constraints, and workload characteristics can substantially restrict the attainable reductions \cite{10.1145/3627703.3650079}. Collectively, these studies establish execution timing as a useful but constraint-dependent mechanism for reducing operational computing emissions.

Existing carbon-aware workload-management approaches exploit temporal flexibility at different levels of the computing stack. At the application level, CarbonScaler dynamically adjusts the resources allocated to an elastic batch job according to variations in grid carbon intensity \cite{10.1145/3626788}. More recent cluster-level systems coordinate multiple jobs and resource constraints: GREEN integrates carbon-aware temporal shifting into an ML cluster scheduler while balancing carbon reduction against job completion time \cite{3767955.3768008}, whereas CarbonFlex jointly provisions cluster resources and schedules parallel cloud jobs using cluster-level workload information \cite{hanafy2025carbonflexenablingcarbonawareprovisioning}. Such approaches demonstrate the value of coordinated carbon-aware management, but generally assume integration with the underlying scheduler or control over resource allocation and provisioning. CATS takes a complementary, lightweight approach by using a proposed job's duration and forecast carbon intensity to recommend a lower-carbon start time that a user can apply directly or pass to an existing queueing system \cite{Bartholomew2025}. This user-facing advisory model enables carbon-aware timing without requiring replacement of the underlying scheduler and directly inspires the external design adopted by COAST.

However, the advisory setting introduces a coordination challenge that is different from the capacity-management problem addressed by internal cluster schedulers. CATS recommends a favorable execution window for a proposed task based on its duration and the forecast carbon intensity profile, but it does not explicitly coordinate that recommendation with those issued to other tasks \cite{Bartholomew2025}. When multiple flexible jobs consult the same forecast over a common planning horizon, independently selected low-carbon windows may therefore coincide. Moreover, because an HPC job occupies resources throughout its runtime, jobs assigned different start times may still overlap across several execution slots. The resulting temporal concentration is an externality: the desirability of a candidate start time for one job depends partly on the execution intervals selected by other jobs. An external advisory layer cannot resolve this interaction through direct queue control or resource provisioning, and must instead account for it while generating the recommendations themselves. This motivates a mechanism that coordinates batches of timing requests while explicitly representing each job's marginal contribution to the future execution-load profile.

Congestion games provide a natural framework for this form of coordination because they model decisions in which multiple participants select shared resources and each participant's cost depends on the resulting resource load \cite{10.1007/BF01737559}. In the COAST setting, jobs act as players, feasible target start times define their strategies, and the execution slots occupied over each job's runtime form the shared temporal resources. This formulation captures the fact that a timing decision affects not only the selecting job's carbon emissions and waiting time, but also the load experienced across the slots in which it executes. Congestion games also admit potential-function representations that align unilateral cost improvements with decreases in a global scalar function \cite{MONDERER1996124}. This property enables decentralized best-response updates to reach a stable assignment in a finite game. COAST applies this structure to extend carbon-aware timing advice from independent recommendations to congestion-aware coordination across micro-batches of flexible HPC jobs.

\section{Methodology}
\label{sec:method}

\subsection{System Model and Assumptions}
\label{sec:system-model}

COAST is an external timing-advice layer for flexible HPC workloads.
Users submit job intents through a local plugin, while the underlying
HPC scheduler remains unchanged. The platform operates in micro-batches of length
\(\Delta=5\) minutes. At decision round \(k\), let
\(\mathcal{J}_k\) denote the set of job intents received during the
current micro-batch. Time is discretized into slots of length
\(\delta=30\) minutes, indexed on a single global timeline shared
across rounds. COAST considers a future horizon of \(H=48\) hours; at
round \(k\), it plans over the rolling set \(\mathcal{T}_k\) of the

\begin{equation}
M = H/\delta = 96
\label{eq:time-horizon}
\end{equation}

consecutive slots that begin at the round's decision time. All slot
indices \(a_i, t, D_i\) below refer to this global timeline.

We set \(\delta=30\) minutes to match the native 30-minute settlement
resolution of the grid carbon intensity trace; a finer grid would
introduce artificial temporal resolution.

Each job \(i\in\mathcal{J}_k\) is characterized by
\begin{equation}
(a_i, r_i, d_i, \hat e_i, D_i, m_i),
\label{eq:job-tuple}
\end{equation}
where \(a_i\) is the arrival slot,
\(r_i\) is the resource demand,
\(d_i\) is the estimated runtime,
\(\hat e_i\) is the estimated total energy,
\(D_i\) is the deadline slot, and \(m_i\) is the maximum number of
slots by which the start may be delayed. Slot indices
(\(a_i, t, D_i\)) and slot counts (\(\ell_i, m_i\)) are on the grid of
\eqref{eq:time-horizon}; \(\delta\) converts a slot difference to
physical time only in the delay cost below. In our trace-driven
evaluation, the arrival \(a_i\), the
resource demand \(r_i\) (from CPU allocation, \(\texttt{AllocCPUS}\)),
and the runtime \(d_i\) are taken directly from the workload trace.
Since the trace does not record per-job energy, the estimated energy
is derived from a linear power model of the allocated resources,

\begin{equation}
\hat e_i = \pi\, r_i\, d_i,
\label{eq:energy-model}
\end{equation}

where \(\pi\) is a per-core power coefficient, so that power is
proportional to the allocated cores. The deadline \(D_i\) and the
maximum delay \(m_i\) are likewise absent from the trace; we synthesize
them from the runtime as \(m_i=\min(\lceil 4 d_i/\delta\rceil,\, 48)\)
slots (up to four times the runtime, capped at 24 h) and
\(D_i = a_i + m_i + \lceil d_i/\delta\rceil\). With this construction
the finish-by deadline coincides with the delay bound, so the deadline
term in \eqref{eq:feasible-start} does not bind independently in our
experiments.

For notational convenience, let
\begin{equation}
\ell_i = \left\lceil d_i/\delta \right\rceil
\label{eq:ell}
\end{equation}
denote the number of execution slots occupied by job \(i\).

For algorithmic analysis and trace-driven evaluation, we assume that an
accepted recommendation is realized at its target execution start time.
This isolates the carbon--waiting--congestion trade-off from the
queueing behavior of the external HPC scheduler. Predicting queue delay
and converting target execution times into practical submission times
are outside the scope of this work.

COAST uses a soft congestion penalty rather than a hard capacity
constraint. This is appropriate for an external advisory platform,
which does not have access to the full capacity, queue state, or
priority policy of the underlying HPC system.

\subsection{Problem Formulation}
\label{sec:problem-formulation}

A job selects one feasible target execution start time. The feasible
start-time set of job \(i\) is

\begin{equation}
\mathcal{S}_i =
\left\{
t\in\mathcal{T}_k \;\middle|\;
t \geq a_i,\;
t-a_i \leq m_i,\;
t+\ell_i-1 \leq D_i
\right\}.
\label{eq:feasible-start}
\end{equation}

Membership \(t\in\mathcal{T}_k\) already confines starts to the
48-hour forward horizon, so no separate horizon constraint is needed.

A strategy profile is denoted by
\begin{equation}
\bm{s}=(s_i)_{i\in\mathcal{J}_k},
\qquad
s_i\in\mathcal{S}_i.
\label{eq:strategy-profile}
\end{equation}

Let \(c_\tau\) be the predicted grid carbon intensity at execution
slot \(\tau\), expressed in \(\mathrm{gCO_2/kWh}\). We assume that the
estimated job energy is distributed uniformly across its estimated
runtime. The predicted carbon emission of job \(i\), if it starts at
slot \(t\), is therefore

\begin{equation}
\mathcal{C}_i(t)
=
\sum_{\tau=t}^{t+\ell_i-1}
\frac{\hat e_i}{\ell_i}c_\tau .
\label{eq:carbon-cost}
\end{equation}

This form allows COAST to use job-level energy observations directly in
trace-driven evaluation. More detailed power profiles, when available,
can replace the uniform-energy approximation without changing the
game formulation.

The waiting cost is proportional to the delay between the job arrival
and its target execution start time:

\begin{equation}
\mathcal{W}_i(t)
=
\delta(t-a_i).
\label{eq:waiting-cost}
\end{equation}

At the beginning of round \(k\), previous accepted recommendations
contribute a background load \(B_\tau^k\) at each execution slot
\(\tau\). Given a strategy profile \(\bm{s}\), the total execution-slot
load is

\begin{equation}
L_\tau(\bm{s})
=
B_\tau^k
+
\sum_{i\in\mathcal{J}_k}
r_i
\mathbf{1}
\left\{
s_i \leq \tau < s_i+\ell_i
\right\},
\label{eq:execution-load}
\end{equation}

where \(\mathbf{1}\{\cdot\}\) is the indicator function. Unlike a
start-slot-only model, \eqref{eq:execution-load} captures the
overlapping advised resource demand throughout the full runtime of
jobs.

\subsection{Micro-Batch Congestion Game}
\label{sec:congestion-game}

For each micro-batch \(k\), COAST defines the game

\begin{equation}
\mathcal{G}_k =
\left(
\mathcal{J}_k,
\{\mathcal{S}_i\}_{i\in\mathcal{J}_k},
\{J_i\}_{i\in\mathcal{J}_k}
\right).
\label{eq:game}
\end{equation}

Each job is modeled as a player, and its strategy is a feasible target
start time \(s_i\in\mathcal{S}_i\). Although the platform centrally
computes the recommendations, this game-theoretic abstraction captures
the fact that the desirability of a start time depends on the choices
of other jobs in the same batch.

Let the load at slot \(\tau\), excluding job \(i\), be

\begin{equation}
L_\tau^{-i}(\bm{s})
=
B_\tau^k
+
\sum_{\substack{j\in\mathcal{J}_k\\j\neq i}}
r_j
\mathbf{1}
\left\{
s_j \leq \tau < s_j+\ell_j
\right\}.
\label{eq:load-excluding-i}
\end{equation}

We define the slot-level congestion potential as

\begin{equation}
\psi(x)=x^2.
\label{eq:congestion-potential}
\end{equation}

If job \(i\) chooses start time \(t\), its marginal congestion
increment is the cumulative increase in congestion potential over all
slots occupied by the job:

\begin{equation}
\Delta_i(t,\bm{s}_{-i})
=
\sum_{\tau=t}^{t+\ell_i-1}
\left[
\psi\!\left(L_\tau^{-i}(\bm{s})+r_i\right)
-
\psi\!\left(L_\tau^{-i}(\bm{s})\right)
\right].
\label{eq:marginal-congestion}
\end{equation}

The cost of job \(i\) is then

\begin{equation}
J_i(t,\bm{s}_{-i})
=
\alpha \frac{\mathcal{C}_i(t)}{\hat c}
+
\beta \frac{\mathcal{W}_i(t)}{\hat w}
+
\gamma \frac{\Delta_i(t,\bm{s}_{-i})}{\hat g},
\label{eq:job-cost}
\end{equation}

where \(\alpha,\beta,\gamma\geq0\) determine the relative importance of
carbon emissions, waiting time, and congestion. The constants
\(\hat c,\hat w,\hat g>0\) are fixed normalization constants (medians
of the respective raw terms, computed once on training data and then
frozen) that place the three components on comparable scales. Because
they are constants, they can be folded into \(\alpha,\beta,\gamma\) and
do not affect the exact-potential property established below. Thus,
COAST recommends start times that are low-carbon, acceptable in delay,
and unlikely to create concentrated execution load.

\subsection{COAST Algorithm and Theoretical Property}
\label{sec:coast-algorithm}

COAST computes a stable assignment for each micro-batch using
sequential strict best-response updates. It first initializes each job
with any feasible start time, for example the feasible slot with the
lowest carbon-plus-waiting cost under background load \(B_\tau^k\).
It then repeatedly visits jobs and moves a job only when another
feasible start time strictly decreases its cost in
\eqref{eq:job-cost}. Ties are resolved by retaining the current
start time.

\begin{algorithm}
\caption{COAST micro-batch coordination}
\label{alg:coast}
\begin{algorithmic}[1]
\STATE \textbf{Input:} batch \(\mathcal{J}_k\), feasible sets
\(\{\mathcal{S}_i\}\), carbon trace \(\{c_\tau\}\), background load
\(\{B_\tau^k\}\)
\STATE \textbf{Initialize:} assign each \(i\in\mathcal{J}_k\) a feasible
start time \(s_i\in\mathcal{S}_i\)
\REPEAT
    \STATE \(\textit{updated} \gets \textbf{false}\)
    \FOR{each \(i\in\mathcal{J}_k\)}
        \STATE \(t^\star \gets
        \arg\min_{t\in\mathcal{S}_i} J_i(t,\bm{s}_{-i})\)
        \IF{\(J_i(t^\star,\bm{s}_{-i}) <
        J_i(s_i,\bm{s}_{-i})\)}
            \STATE \(s_i \gets t^\star\)
            \STATE \(\textit{updated} \gets \textbf{true}\)
        \ENDIF
    \ENDFOR
\UNTIL{\(\textit{updated}=\textbf{false}\)}
\STATE \textbf{Output:} recommended start times \(\bm{s}\)
\end{algorithmic}
\end{algorithm}

A game is an exact potential game~\cite{MONDERER1996124} if there is a
function \(\Phi\) such that, for every job \(i\), every
\(\bm{s}_{-i}\), and every \(a,b\in\mathcal{S}_i\),

\begin{equation}
J_i(b,\bm{s}_{-i}) - J_i(a,\bm{s}_{-i})
=
\Phi(b,\bm{s}_{-i}) - \Phi(a,\bm{s}_{-i}).
\label{eq:exact-potential}
\end{equation}

For \(\mathcal{G}_k\) we use the batch-level potential

\begin{equation}
\Phi(\bm{s})
=
\sum_{i\in\mathcal{J}_k}
\left[
\alpha \frac{\mathcal{C}_i(s_i)}{\hat c}
+
\beta \frac{\mathcal{W}_i(s_i)}{\hat w}
\right]
+
\frac{\gamma}{\hat g}
\sum_{\tau\in\mathcal{T}_k}
\psi\!\left(L_\tau(\bm{s})\right).
\label{eq:potential-function}
\end{equation}

\textbf{Proposition 1.}
For every fixed micro-batch \(\mathcal{J}_k\), with the background load
\(B^k\) held fixed, the game \(\mathcal{G}_k\) is an exact potential
game with potential \(\Phi(\bm{s})\).

\textit{Proof sketch.}
Consider a unilateral move by job \(i\) from start time \(a\) to
start time \(b\). The carbon and waiting terms in
\eqref{eq:job-cost} change by exactly the same amount as their
corresponding terms in \eqref{eq:potential-function}. For the
congestion term, only execution slots occupied before or after the
move are affected. By \eqref{eq:marginal-congestion}, the change in
job \(i\)'s congestion cost is exactly the change in the summed
quadratic congestion potential. Therefore, the change in \(J_i\)
equals the change in \(\Phi\), establishing the exact potential
property.

The property holds because each job's congestion cost is defined as
its marginal contribution to the global load penalty,
\(\Delta_i(t,\bm{s}_{-i}) = \sum_{\tau}\psi(L_\tau(\bm{s})) -
\sum_{\tau}\psi(L_\tau(\bm{s}_{-i}))\), where the second term is
independent of job \(i\)'s own choice. This construction yields an
exact potential for any convex \(\psi\) and any heterogeneous resource
demands \(r_i\); it does not rely on the affine special case for which
weighted congestion games are otherwise known to admit an exact
potential.

\textbf{Corollary 1.}
Sequential strict best-response updates in Algorithm~\ref{alg:coast}
converge in finitely many steps to a batch-level pure-strategy Nash
equilibrium.

\textit{Justification.}
Each successful update strictly decreases the exact potential
\(\Phi\). Since every job has a finite feasible strategy set and the
batch contains finitely many jobs, the joint strategy space is finite,
so the potential cannot decrease indefinitely and the algorithm
terminates at a profile where no job can improve by a unilateral change
of start time. This equilibrium is defined within a single micro-batch
and its fixed background load; the rolling procedure applies
Algorithm~\ref{alg:coast} to each batch in turn on the accumulating
committed load, and makes no claim of a global optimum or a cross-batch
equilibrium.

Two properties make this game view useful beyond convergence. First,
computing the global minimizer of \(\Phi\) is NP-hard in general (it
subsumes balancing load to minimize the sum of squared slot loads), so
the exact-potential structure is what lets cheap best-response serve as
a principled solver: convergence is guaranteed and each update is a
one-dimensional \(\arg\min\), though the equilibrium is a local rather
than global minimum of \(\Phi\). Second, the per-job cost \(J_i\) keeps
each recommendation individually acceptable, since no job is moved to a
slot that raises its own cost; this suits an advisory layer that users
may decline and admits a decentralized deployment in which jobs respond
on their own.

The resulting recommendations are provisional until users respond.
Let \(y_i^k\in\{0,1\}\) indicate whether job \(i\) accepts its
recommendation in round \(k\). The accepted background load used in
the next round is updated as

\begin{equation}
B_\tau^{k+1}
=
\widetilde{B}_\tau^k
+
\sum_{i\in\mathcal{J}_k}
y_i^k r_i
\mathbf{1}
\left\{
s_i \leq \tau < s_i+\ell_i
\right\},
\label{eq:background-update}
\end{equation}

where \(\widetilde{B}_\tau^k\) denotes the previous accepted load after
the horizon is advanced and expired commitments are removed. A rejected
or timed-out recommendation has \(y_i^k=0\) and does not contribute to
future background load.

\section{Evaluation}

\subsection{Experimental Setup}
\label{sec:setup}

\textbf{Workload trace.}
We drive the simulation with the public CEA-Curie 2011 log from the
Parallel Workloads Archive~\cite{FEITELSON20142967,curie_swf}, an
archive-cleaned trace of a large production HPC system. Each job
supplies its submission time (\(a_i\)), allocated cores
(\(\texttt{AllocCPUS}\), \(r_i\)), and runtime (\(d_i\)). The trace
records no per-job energy, deadline, or delay tolerance, so the energy
\(\hat e_i\) (linear power model, \(\pi=10\,\mathrm{W}\)/core), deadline
\(D_i\), and maximum delay \(m_i\) follow
Section~\ref{sec:system-model}. We keep jobs with runtime
\(\geq 60\,\mathrm{s}\), cap cores at 4096, and discretize onto the
30-minute grid, leaving 165{,}887 jobs across 224 day-long windows
(median 629 jobs/window). Jobs are heterogeneous: median runtime
18\,min (90th percentile 11.7\,h) and 64 cores (90th percentile 1024),
spanning orders of magnitude.

\textbf{Carbon intensity trace.}
Grid carbon intensity is the public UK Carbon Intensity API (National
Grid ESO)~\cite{ukci_api}: 17{,}557 half-hourly \(\mathrm{gCO_2/kWh}\)
samples over 1 January 2025 to 1 January 2026 (range 24--293, median
124). While the HPC was not located in the UK, we have used UK carbon intensity data due to the absence of historical carbon intensity data specific to the HPC dataset used. As the 2011 workload
predates it, we pair counterfactually: each window draws a random valid
start index into the trace and aligns with the carbon data beginning
there, not a calendar replay.

\textbf{Simulation and baselines.}
Evaluation is rolling-horizon: 5-minute micro-batches are solved over a
fresh 48-hour horizon against previously committed load on a shared
timeline; every accepted recommendation is honored and scored at the
actual carbon intensity of its execution window
(Section~\ref{sec:system-model}). Three baselines share the feasibility
model and attributes: Observed (no shifting, earliest feasible slot),
Carbon-greedy (lowest-carbon slot), and Carbon-plus-waiting (weighted
carbon and waiting); all optimize each job in isolation (\(\gamma=0\)),
while only COAST couples jobs.

\textbf{Metrics and split.}
We report carbon, average added waiting (hours), peak slot load, and
load variance; except waiting, each is a signed percentage change
versus Observed, and the reported 95\% intervals are
normal-approximation (mean \(\pm\,1.96\) standard errors). The 224
windows split at random (fixed permutation, seed 0; not chronological)
into disjoint 67 training and 157 held-out test windows
(\(\approx\)30/70); the weights \((\alpha,\gamma)\) and normalizers
\(\hat c,\hat w,\hat g\) are calibrated on training only and frozen,
preventing test leakage.

\subsection{Evaluation Results}
\label{sec:results}

\textbf{Weight selection and operating point.}
The three cost weights and three normalization constants are fixed on
the 67 training windows and then frozen; no held-out window influences
them. We set \(\beta=1\) as the reference unit, fix \(\hat w=60\)
minutes, and take \(\hat c\) and \(\hat g\) as the medians of the raw
carbon and congestion terms over a training calibration pool, so the
three terms share a comparable scale. A median is robust to the
heavy-tailed costs, and scaling rather than centering (as in
standardization) keeps the costs non-negative and preserves the
exact-potential property. We then grid-search
\(\alpha\in\{1,2,4\}\) and ten values of \(\gamma\) from
\(2{\times}10^{-4}\) to \(1\): with the terms already normalized,
\(\alpha\) needs only an \(O(1)\) range, while \(\gamma\) is swept
logarithmically because its scale is not known a priori and the useful
settings are small, where congestion acts as a mild regularizer.

The operating point is chosen on the training windows by maximizing
peak-load reduction subject to retaining at least 85\% of
carbon-greedy's carbon saving (a reduction of at least 12.6\%) and
average added waiting below three hours. The frozen configuration is
\(\alpha=4\), \(\beta=1\), \(\gamma=5{\times}10^{-4}\)
(Table~\ref{tab:config}), used for all held-out results below. Larger
\(\gamma\) trades a little carbon saving and waiting for stronger peak
suppression, an aspect we examine on held-out data below.

\begin{table}[t]
\centering
\caption{Frozen COAST configuration, calibrated on the 67 training
windows and held fixed for all held-out results.}
\label{tab:config}
\begin{tabular}{l c}
\toprule
Component & Value \\
\midrule
Carbon-weight search grid \(\alpha\) & \(\{1,\,2,\,4\}\) \\
Congestion-weight grid \(\gamma\) & \(\{2{\times}10^{-4},\ldots,1\}\) (10 values) \\
Waiting weight \(\beta\) (fixed) & \(1\) \\
\midrule
Selected carbon weight \(\alpha\) & \(4\) \\
Selected congestion weight \(\gamma\) & \(5{\times}10^{-4}\) \\
Carbon normalizer \(\hat c\) & \(82.66\) \\
Waiting normalizer \(\hat w\) & \(60\) min \\
Congestion normalizer \(\hat g\) & \(9216\) \\
\bottomrule
\end{tabular}
\end{table}

\begin{figure*}[t]
\centering
\includegraphics[width=0.8\textwidth]{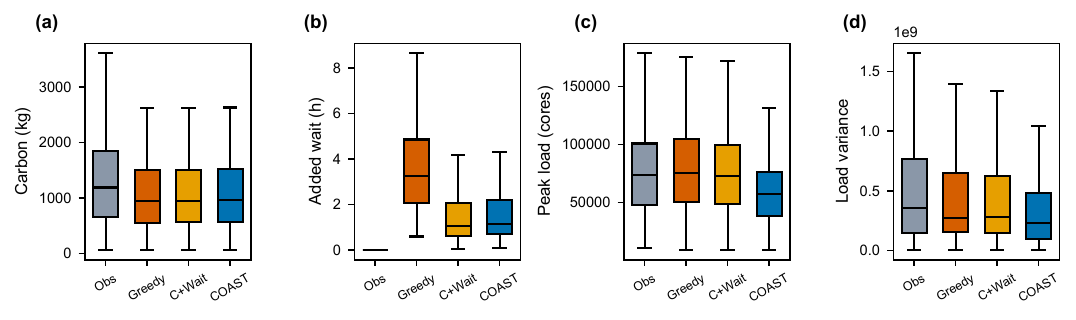}
\caption{Distributions over the 157 held-out test windows: (a) absolute
carbon emissions, (b) additional waiting, (c) peak execution-slot load, and
(d) load variance. Boxes omit fliers for readability. COAST matches the carbon-focused baselines
on carbon while achieving the lowest peak load and load variance.}
\label{fig:metrics_box}
\end{figure*}

\textbf{Carbon--waiting--load trade-off.} Table~\ref{tab:main} and Fig.~\ref{fig:metrics_box} report COAST cuts carbon by 15.4\%
versus Observed (95\% CI \([13.7,17.2]\)), within about a point of the
two carbon-focused baselines (whose CIs overlap) and retaining 92.8\%
of carbon-greedy's saving, at 1.5 h of added waiting against 1.4 h for
carbon-plus-waiting and 3.6 h for carbon-greedy.

The decisive difference is peak load. Both carbon-focused baselines
raise peak execution-slot load (carbon-greedy by 8.5\%,
carbon-plus-waiting by 5.2\%) by concentrating jobs into the same
slots, whereas COAST lowers it: by 20.2\% versus Observed and 23.8\%
versus carbon-greedy, with the lowest load variance (about 35\% below
Observed). Paired Wilcoxon signed-rank tests over the 157 windows
(Table~\ref{tab:sig}) confirm that COAST's carbon reduction versus
Observed and its peak reductions versus Observed, carbon-greedy, and
carbon-plus-waiting are all significant at \(p<10^{-4}\) and hold in
most windows. The cost is a significant increase in waiting (median
\(+1.15\) h) and a carbon saving about one point short of carbon-greedy
(median \(+1.12\%\) more carbon), so congestion awareness sacrifices
almost none of the carbon benefit.

\begin{table}[t]
\centering
\caption{Paired Wilcoxon signed-rank tests over the 157 held-out
windows (two-sided): COAST versus each baseline.}
\label{tab:sig}
\begin{tabular}{l l r r c}
\toprule
Baseline & Metric & Median & Lower & \(p\) \\
\midrule
Observed        & Carbon  & \(-13.19\%\)   & 154/157 & \(<10^{-4}\) \\
                & Peak    & \(-19.94\%\)   & 135/157 & \(<10^{-4}\) \\
                & Waiting & \(+1.15\)\,h    & 0/157   & \(<10^{-4}\) \\
\addlinespace
Carbon-greedy   & Peak    & \(-23.93\%\)   & 154/157 & \(<10^{-4}\) \\
                & Carbon  & \(+1.12\%\)    & 0/157   & \(<10^{-4}\) \\
\addlinespace
Carbon+waiting  & Peak    & \(-20.64\%\)   & 147/157 & \(<10^{-4}\) \\
\bottomrule
\end{tabular}
\end{table}

\begin{table*}[t]
\setlength{\tabcolsep}{12pt}
\centering
\caption{Per-window means over the 157 held-out test windows;
\(\pm\) is the 95\% CI half-width. \(\Delta\) is method minus Observed,
and \(\downarrow\) marks that lower is better. Weights frozen at
\(\alpha=4\), \(\beta=1\),
\(\gamma=5{\times}10^{-4}\).}
\label{tab:main}
\renewcommand{\arraystretch}{1.15}
\begin{tabular}{l r r r r r}

\toprule
& \multicolumn{2}{c}{Absolute (per-window mean)} & \multicolumn{3}{c}{Change vs Observed} \\
\cmidrule(lr){2-3}\cmidrule(lr){4-6}
Method & Carbon (kg) & Peak (cores) & \(\Delta\)Carbon (\%)~\(\downarrow\) & \(\Delta\)Peak (\%)~\(\downarrow\) & \(\Delta\)Waiting (h)~\(\downarrow\) \\
\midrule
Observed
  & \(1346.6541\) & \(79232.5605\)
  & \(0.0000 \pm 0.0000\) & \(0.0000 \pm 0.0000\) & \(0.0000 \pm 0.0000\) \\
Carbon-greedy
  & \(1116.0222\) & \(83845.0892\)
  & \(-16.6097 \pm 1.7500\) & \(+8.4694 \pm 5.6018\) & \(+3.6037 \pm 0.3156\) \\
Carbon+waiting
  & \(1122.8616\) & \(81567.8981\)
  & \(-15.9557 \pm 1.7203\) & \(+5.1601 \pm 5.3335\) & \(+1.4365 \pm 0.1826\) \\
COAST
  & \(1130.6650\) & \(59769.1529\)
  & \(-15.4057 \pm 1.7453\) & \(-20.1804 \pm 3.5777\) & \(+1.5289 \pm 0.1824\) \\
\bottomrule
\end{tabular}
\end{table*}

\begin{figure*}[t]
\centering
\includegraphics[width=0.8\textwidth]{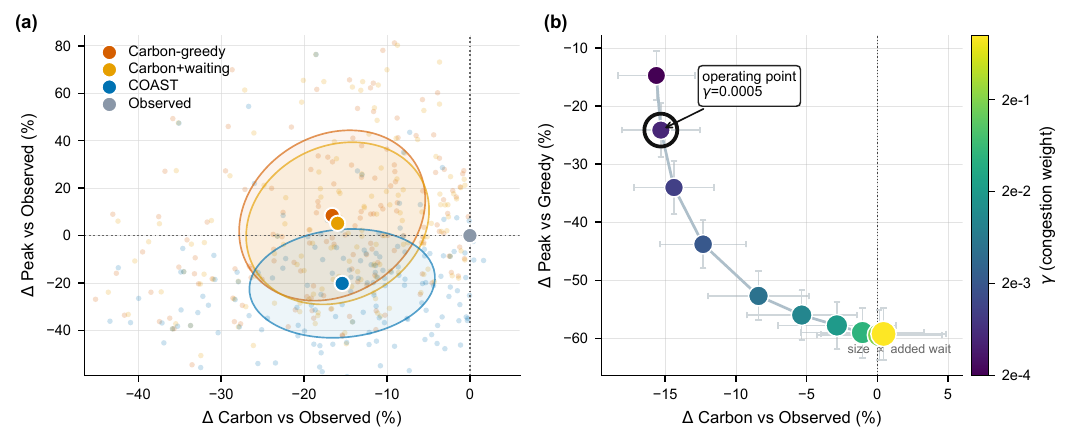}
\caption{(a)~All policies in the signed carbon--peak plane over the
157 held-out windows (percentage change vs Observed; lower-left is
better). (b)~Effect of \(\gamma\) on 40 held-out frontier windows:
carbon reduction vs Observed against peak change vs carbon-greedy,
marker size \(\propto\) added waiting; the frozen point
\(\gamma=5{\times}10^{-4}\) (circled) was selected on training.}
\label{fig:frontier}
\end{figure*}

\textbf{Effect of congestion awareness.} Fig.~\ref{fig:frontier}(a) shows that the carbon-focused baselines reduce
carbon emissions at the cost of higher peak load because independently
chasing the cleanest slots concentrates jobs. In contrast, only COAST
reduces both metrics, and the scatter and
covariance ellipses indicate that this improvement is consistent across
windows.

Fig.~\ref{fig:frontier}(b) shows the effect of the congestion weight
\(\gamma\). Increasing \(\gamma\) yields greater peak reduction with
only a small loss in carbon reduction and a modest increase in waiting.
The selected value, \(\gamma=5{\times}10^{-4}\), lies near the knee of
the trade-off curve.

\begin{figure*}[t]
\centering
\includegraphics[width=0.8\textwidth]{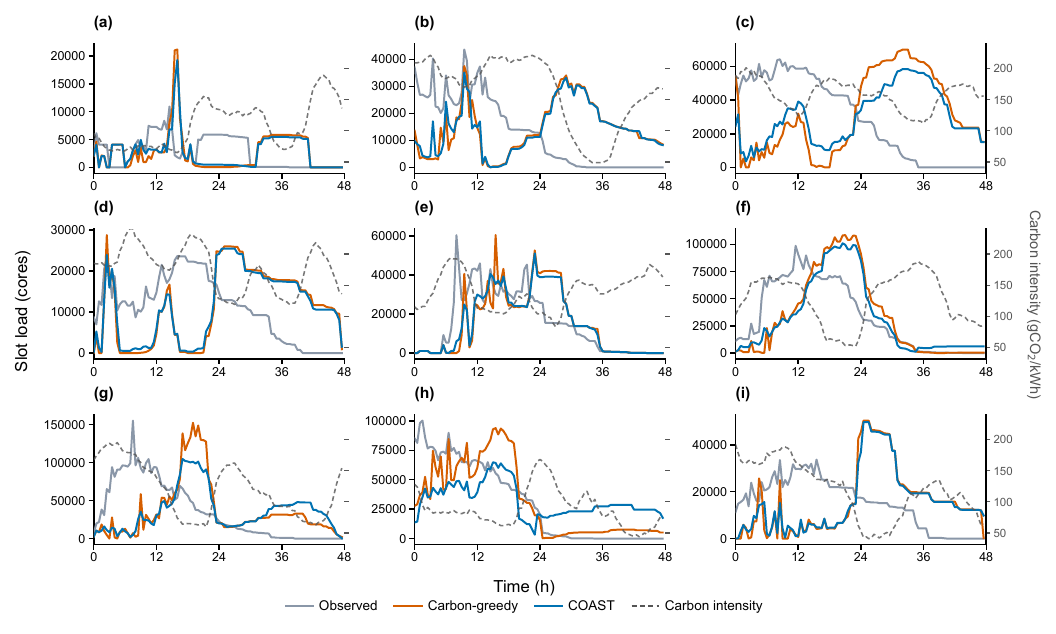}
\caption{First nine held-out windows, each showing the 48-hour interval
of highest carbon-greedy load: Observed, carbon-greedy, and COAST slot
load (left axis) and carbon intensity (right axis, dashed). Panels
illustrate the mechanism; quantitative results are in
Table~\ref{tab:main}.}
\label{fig:profiles}
\end{figure*}

\textbf{Reshaping load over time.}
Fig.~\ref{fig:profiles} shows how congestion awareness reshapes load over time. Carbon-greedy creates tall, narrow spikes because flexible jobs independently converge on the same low-carbon periods. COAST
instead spreads jobs across nearby feasible slots, preserving most of
the carbon benefit while substantially lowering peak load. This
time-domain smoothing explains the reductions in aggregate peak load
and load variance reported in Table~\ref{tab:main}.

\begin{figure*}[t]
\centering
\includegraphics[width=0.8\textwidth]{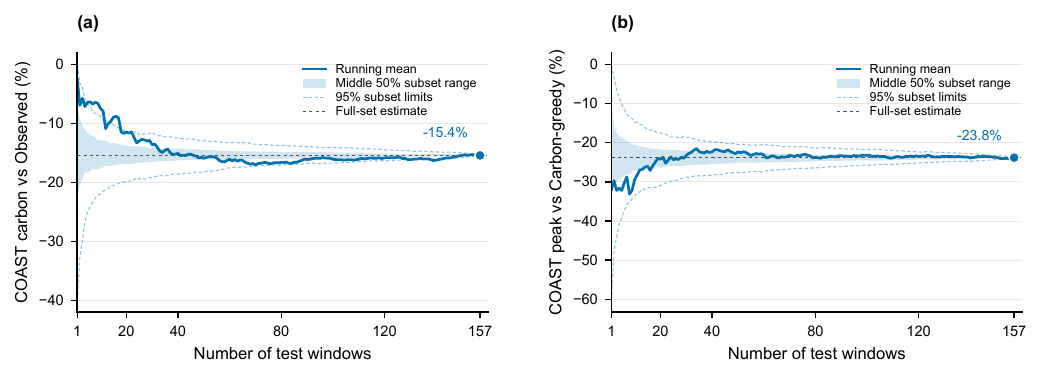}
\caption{Headline estimates as test windows accumulate:
(a)~COAST carbon vs Observed, (b)~COAST peak vs carbon-greedy. Solid
line: running mean under one fixed ordering; band and dashed lines:
middle 50\% and 95\% ranges over 500 random permutations of the 157
windows; horizontal line: full-set estimate.}
\label{fig:convergence}
\end{figure*}

\textbf{Stability across test windows.}
Fig.~\ref{fig:convergence} shows that the estimated carbon and peak
reductions converge rapidly to their full-set values as more test
windows are included. The narrowing uncertainty bands indicate that
the estimates become increasingly stable, demonstrating that the
reported results are robust across the 157 held-out windows.

\begin{figure*}[t]
\centering
\includegraphics[width=0.8 \textwidth]{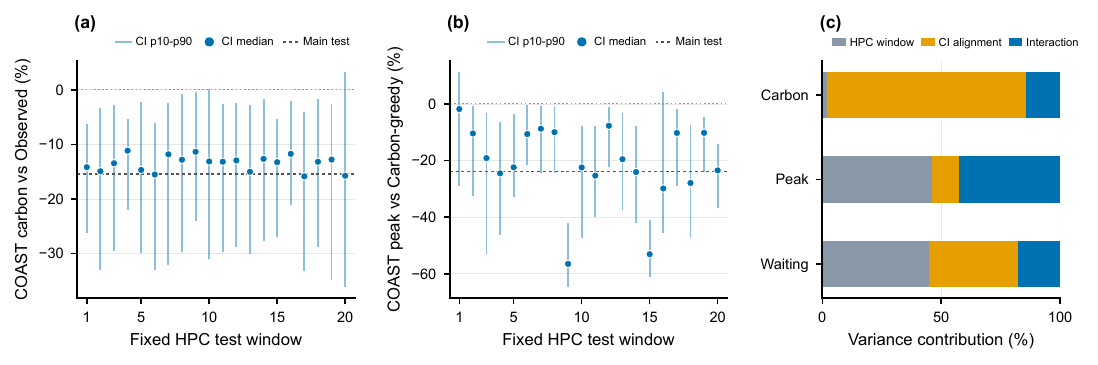}
\caption{Crossed carbon-alignment sensitivity (20 HPC windows
\(\times\) 20 alignments). (a,b)~For each fixed HPC window, the
p10--p90 range and median across the 20 carbon alignments of COAST's
carbon reduction vs Observed and peak reduction vs carbon-greedy;
dashed line is the main test value. (c)~Two-factor variance
decomposition (HPC window, carbon alignment, interaction) per metric.}
\label{fig:sens}
\end{figure*}

\textbf{Sensitivity to carbon alignment.}
We evaluate 20 held-out HPC windows under 20 carbon alignments (400
combinations), with all parameters fixed (Table~\ref{tab:sens},
Fig.~\ref{fig:sens}). The crossed averages remain close to the main
results, indicating that the reported improvements are not sensitive to
a particular workload--carbon pairing.

Variance decomposition further shows that carbon reduction is dominated
by carbon alignment (83\% of the variance), whereas peak reduction is
mainly determined by workload structure and its interaction with carbon
alignment (46\% and 42\%, respectively). Waiting time depends on both
factors.

\begin{table}[t]
\centering
\caption{Crossed carbon-alignment sensitivity. Crossed denotes the mean over 400 pairs; Main denotes the 157-window test results.}
\label{tab:sens}
\renewcommand{\arraystretch}{1.15}
\begin{tabular}{l r r r r r}
\toprule
& & & \multicolumn{3}{c}{Variance share (\%)} \\
\cmidrule(lr){4-6}
Metric & Crossed & Main & HPC & CI & Int. \\
\midrule
Carbon vs Obs (\%)   & \(-14.99\) & \(-15.41\) & 2.1  & 83.4 & 14.5 \\
Peak vs Obs (\%)     & \(-19.27\) & \(-20.18\) & 46.1 & 24.2 & 29.6 \\
Peak vs Greedy (\%)  & \(-21.71\) & \(-23.85\) & 46.2 & 11.4 & 42.5 \\
Waiting (h)          & \(1.65\)   & \(1.53\)   & 44.7 & 37.4 & 17.9 \\
\bottomrule
\end{tabular}
\end{table}

\textbf{Algorithm overhead.} Best-response converges almost immediately: across the 25{,}116 non-empty micro-batches in the test set (160 per window), 98\% reach a stable assignment in a single sweep, averaging 1.02 sweeps and at most 7. Each micro-batch solves in a median of 0.07\,ms (mean 0.16\,ms) on a single core of an Apple M3 Pro (performance core up to 4.05\,GHz, 18\,GB RAM), so a full day of recommendations costs about 25\,ms. These are indicative solver times, not an end-to-end runtime benchmark.

\section{Discussion and Limitations}

COAST is evaluated as an advisory layer under deliberately idealized
assumptions. Accepted recommendations are assumed to be realized at
their target slots: we do not model queue delay or translate target
start times into submission times, so the reported gains represent an
upper bound for an external advisor. Candidates are scored using the
actual carbon intensity of their execution window rather than forecasts.
Because the CEA-Curie trace lacks per-job energy, we estimate it with a
linear power model and pair the 2011 workload with UK carbon intensity
under random alignments. Finally, the congestion penalty is soft and
does not model system capacity, queue state, or priority policy. These
assumptions isolate the carbon--waiting--congestion trade-off, leaving
queueing, forecast error, and scheduler enforcement to future work.

\section{Conclusion}

Independently shifting HPC jobs toward low-carbon periods can create
execution-time congestion and new load peaks. COAST addresses this by
coordinating each micro-batch as an exact potential game in which every
job internalizes its marginal contribution to slot congestion,
converging through best-response updates to a stable set of start-time
recommendations. Across a year of grid carbon intensity data and 157
held-out workload days, COAST matches the carbon savings of
congestion-agnostic carbon-aware timing (15.4\% below no shifting)
while being the only policy that reduces rather than raises peak load
(by 20.2\% versus no shifting and 23.8\% versus carbon-greedy), at
modest added waiting. Congestion-aware temporal coordination can thus
capture most of the carbon benefit of workload shifting without the
load concentration it would otherwise cause, a property that grows more
important as more flexible demand is moved onto the same low-carbon
periods.

% \noindent

\bibliographystyle{IEEEtran}
\bibliography{references}

\end{document}